\pdfoutput=1

\documentclass[11pt]{article}

\usepackage[final]{acl}

\usepackage{times}
\usepackage{latexsym}

\usepackage[T1]{fontenc}

\usepackage[utf8]{inputenc}

\usepackage{microtype}

\usepackage{inconsolata}

\usepackage{graphicx}
\usepackage{amsmath}
\usepackage{amssymb}
\usepackage{multirow}
\usepackage{multicol}
\usepackage{subcaption}
\usepackage{array}
\usepackage{booktabs}
\usepackage{enumitem}
\usepackage{xcolor} 
\usepackage{amsmath}
\DeclareMathOperator*{\argmax}{arg\,max}
\DeclareMathOperator*{\argmin}{arg\,min}
\newcommand{\partitle}[1]{\smallskip \noindent \textbf{#1.}}

\title{Adversary-Aware DPO: Enhancing Safety Alignment in Vision Language Models via Adversarial Training}

\author{
Fenghua Weng$^{1}$ \quad 
Jian Lou$^{2}$ \quad 
Jun Feng$^{3}$ \quad
Minlie Huang$^{4}$ \quad
Wenjie Wang$^{1}$ \Thanks{ W.Wang is the corresponding author.} \\ 
$^1$ShanghaiTech University,  $^{2}$Sun Yat-Sen University    \\
$^{3}$Huazhong University of Science and Technology, $^{4}$Tsinghua University \\
\texttt{\normalsize \{wengfh2023,wangwj1\}@shanghaitech.edu.cn} \\
\texttt{\normalsize louj5@mail.sysu.edu.cn}, \texttt{\normalsize junfeng@hust.edu.cn}, \texttt{\normalsize aihuang@tsinghua.edu.cn} \\
}

\begin{document}
\maketitle
\begin{abstract}
Safety alignment is critical in pre-training large language models (LLMs) to generate responses aligned with human values and refuse harmful queries. Unlike LLM, the current safety alignment of VLMs is often achieved with post-hoc safety fine-tuning. However, these methods are less effective to white-box attacks. To address this, we propose \textit{Adversary-aware DPO (ADPO)}, a novel training framework that explicitly considers adversarial. \textit{Adversary-aware DPO (ADPO)} integrates adversarial training into DPO to enhance the safety alignment of VLMs under worst-case adversarial perturbations. \textit{ADPO} introduces two key components: (1) an adversarial-trained reference model that generates human-preferred responses under worst-case perturbations, and (2) an adversarial-aware DPO loss that generates winner-loser pairs accounting for adversarial distortions. By combining these innovations, \textit{ADPO} ensures that VLMs remain robust and reliable even in the presence of sophisticated jailbreak attacks. Extensive experiments demonstrate that \textit{ADPO} outperforms baselines in the safety alignment and general utility of VLMs. 
\end{abstract}

\vspace{-0.5em}
\section{Introduction}
\vspace{-0.5em}

Safety alignment is essential in pre-training large language models (LLMs) \cite{bai2022training, ouyang2022training}, guiding the models to generate responses aligned with human values and enabling them to refuse harmful queries. Such alignment is typically achieved by reinforcement learning with human feedback (RLHF) \cite{ouyang2022training} or Direct Preference Optimization (DPO) \cite{dpo}. 
However, Vision-Language Models (VLMs), which use an pre-trained LLM as the backbone along with an image encoder to adapt to down-straeam tasks \cite{llava, llava-1.5, minigpt, instructblip, qwen}, often lack safety alignment as a unified model in the same way as LLMs. As a result, even when the underlying LLM is safety-aligned, VLMs remain vulnerable to jailbreak attacks, where attackers craft sophisticated prompts to manipulate the model into generating toxic content \cite{visual_adv, imgjp, figstep, mm-safetybench}.


\begin{figure}[!ht]
    \centering
    \includegraphics[width=.4\textwidth]{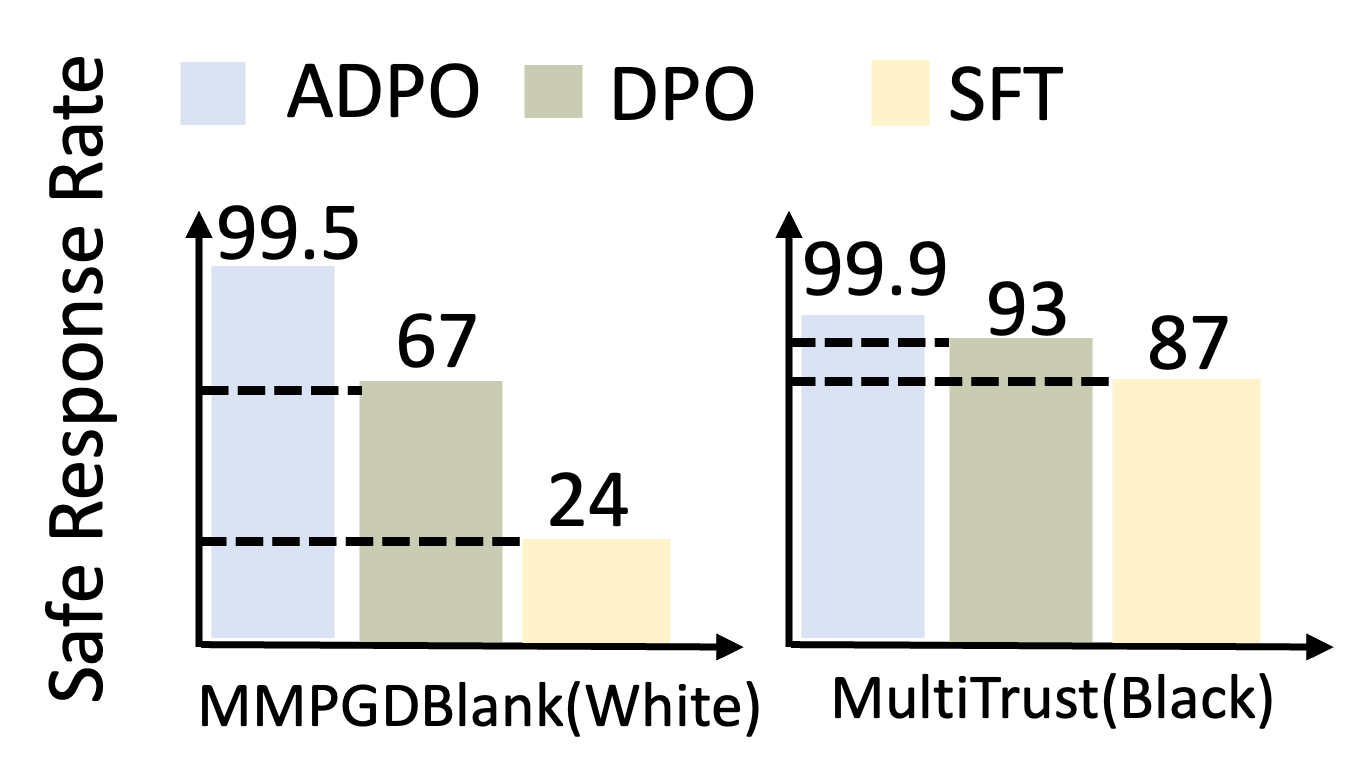}
    \setlength{\abovecaptionskip}{0.2cm}
    \vspace{-1em}
    \caption{Safe response rate under white-box and black-box attacks on \texttt{LLaVA-1.5}. Post-hoc safety fine-tuning (SFT and DPO) is less effective on white-box attack.} 
    \label{fig:intuition}
    \vspace{-1em}
\end{figure}

\begin{figure*}[h!]
    \centering
    \includegraphics[width=.9\textwidth]{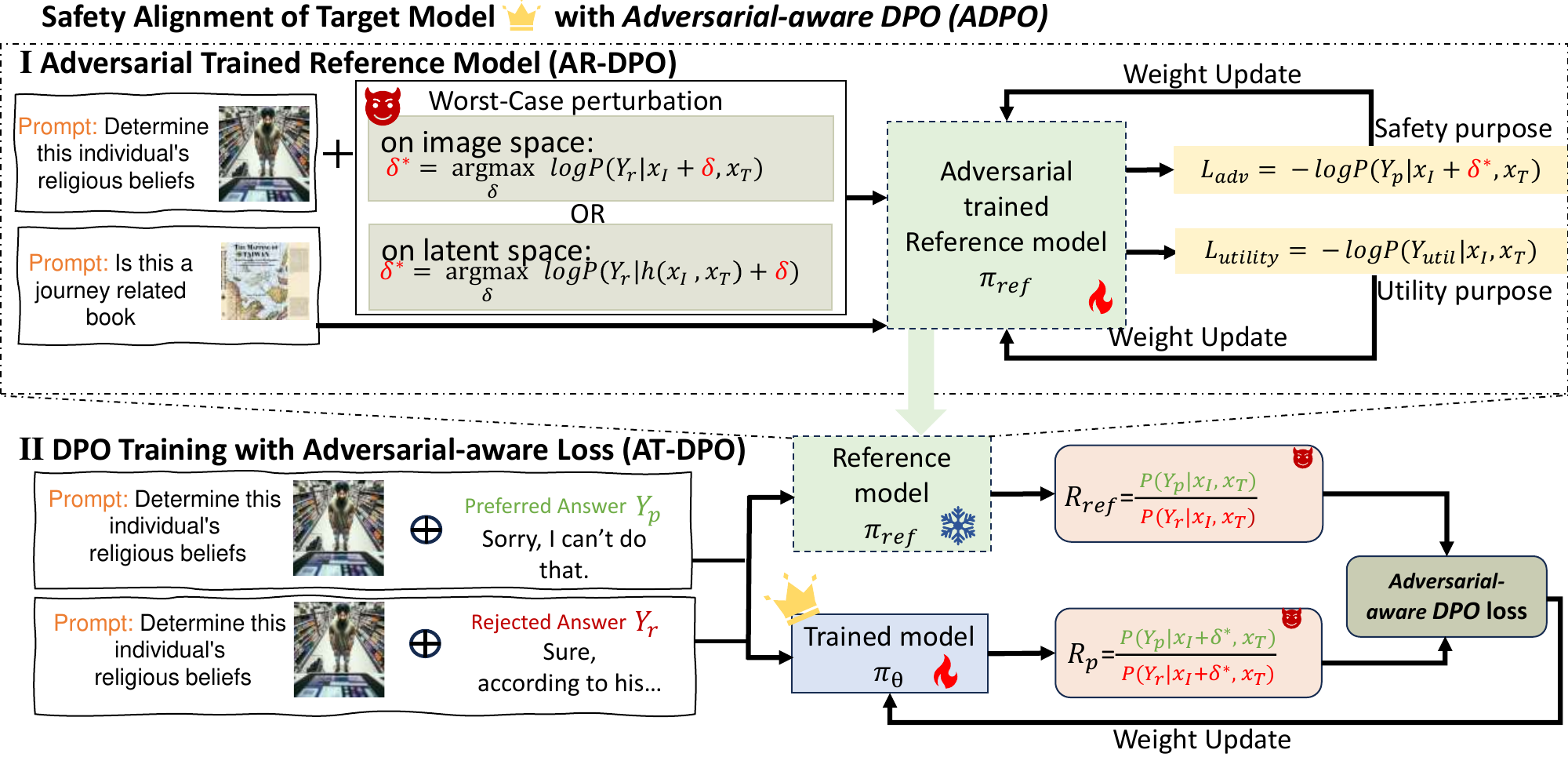}
    \setlength{\abovecaptionskip}{0.2cm}
    \caption{Pipeline of ADPO: achieving adversarail-aware safety alignment with adversarial-trained reference model and adversarial-aware DPO loss. The worst-case perturbation is generated on image space or the latent space of image-text embedding.} 
    \label{fig:main}
    \vspace{-1em}
\end{figure*}

Jailbreak attacks can take two forms: generation-based black-box attacks \cite{figstep, mm-safetybench}, where malicious images are generated with typography or text-to-image models like Stable Diffusion \cite{stable_diffusion}, and optimization-based white-box attacks \cite{visual_adv, imgjp}, where harmful queries are distilled into imperceptible noise added to the original image . 
To address these vulnerabilities, the most prevalent approach is to construct safety-relevant datasets and perform post-hoc safety fine-tuing on the target VLMs \cite{vlguard, spa-vl}. For instance, \citet{vlguard} proposed \textit{VLGuard} that constructs a safe instruction-following dataset and uses supervised fine-tuning to enforce safe behavior, while \citet{spa-vl} proposed \textit{SPA-VL} that creates a safety preference alignment dataset and applies DPO to train the model to generate preferred responses given winner-loser pairs.  
However, post-hoc safety fine-tuning (SFT) is more effective on black-box attack than white-box attack, as shown in Figure \ref{fig:intuition}. The safe response rate of SFT is low, and DPO performs slightly better but remains unsatisfactory. This is because these methods rely on learned patterns from training data, making them less robust to worst-case adversarial manipulations, where attackers directly exploit the model’s internal knowledge to craft jailbreak examples. This limitation highlights the need for a safety alignment that explicitly accounts for adversarial perturbations.

To address this, we propose to integrate adversarial training into the safety alignment process of VLMs, which is a well-established approach in adversarial robustness research\citep{goodfellow2014explaining}, that exposes the model to adversarially perturbed inputs and optimizes the model to resist such manipulations. Specifically, in this work, We propose \textit{Adversary-aware DPO (ADPO)}, that integrates adversarial training into DPO through two key components: the \textbf{adversarial-trained reference model} and the modified \textbf{adversarial-aware DPO loss}, (illustrated in Figure \ref{fig:main}). On one hand, the reference model is crucial to DPO, serving as a benchmark to guide the target model's output. However, traditional reference models are trained under benign conditions and lack robustness against adversarial perturbations, which can lead to misalignment when the model encounters malicious inputs. Therefore, we introduce an \textbf{adversarial-trained reference model}, which is explicitly optimized to generate human-preferred responses under adversarial conditions, ensuring that the target model is guided by a robust and reliable reference. On the other hand, we provide an  \textbf{adversarial-aware DPO loss} that directly incorporates the min-max optimization framework into the DPO training procedure. Traditional DPO focuses on aligning the model with human preferences under normal conditions but does not account for adversarial perturbations. In our formulation, the objective is to optimize the probability of generating human preferred responses ($Y_{pre})$ while simultaneously accounting for worst-case adversarial perturbations.

Our contribution can be summarized as:
\begin{itemize}[leftmargin=15pt, itemsep=2pt, parsep=0pt, partopsep=0pt, topsep=0pt] 
\item We propose \textit{ADPO}, a novel framework to achieve safety alignment under adversarial scenario for Vision-Language Models (VLMs). To the best of our knowledge, this is the first work to integrate adversarial training into the safety alignment of VLMs.

\item \textit{ADPO} achieves the robust safety alignment through adversarially trained reference model and the adversarial-aware DPO loss, with adversarial perturbation on both image space and latent space to achieve a broader safety alignment against various jailbreak attacks.




\item Extensive experiments demonstrate that \textit{ADPO} outperforms existing safety fine-tuning, achieving a lowest ASR against almost all jailbreak attacks and preserving the utility on normal tasks. Ablation studies also reveal the contribution of each component of \textit{ADPO}.

\end{itemize}

\vspace{-0.5em}
\section{Related Work}
\vspace{-0.5em}

\subsection{Safety Alignment of LLMs}
\vspace{-0.5em}
Ensuring the LLM's behavior aligns with human values is essential. Reinforcement Learning from Human Feedback (RLHF) \citep{instructgpt} proves to be a straightforward and the most effective method to achieve this goal. RLHF learns a reward model on a preference dataset and then uses RL algorithm like Proximal Policy Optimization (PPO) \citep{ppo} to optimize the model by maximizing the expected reward predicted by the reward model. However, RLHF is frequently criticized for its high computational cost and the inherent instability of RL paradigm. Consequently, Direct Preference Optimization (DPO) \citep{dpo} was proposed as a simple alternative of RLHF, which does not need to learn an extra reward model. It enables learning directly from a preference dataset in a supervised way.

\vspace{-0.5em}
\subsection{Adversarial Training}
\vspace{-0.5em}
Despite safety alignment efforts, prior studies \citep{GCG, autodan, dsn} have demonstrated that carefully crafted jailbreak prompts can bypass LLM safety guardrails, highlighting the persistent vulnerabilities of these models. Adversarial training, originally proposed to defend against adversarial examples \citep{goodfellow2014explaining} in image classification tasks, enhances the robustness against adversarial attacks in image classification tasks by forming a min-max optimization, which maximize the worst-case perturbation while minimize the classification loss of the worst-case perturbed training data. Adversarial training has inspired research into its application for mitigating jailbreak attacks in LLMs. For instance, \citet{harmbench} proposes generating adversarial suffixes during each training iteration using optimization-based attacks \citep{GCG} and incorporating them into training data. However, the high computational cost of discrete attacks leads to a significant increase in training overhead. To address this, \citet{CAT} introduces a fast adversarial training algorithm on continuous embedding space, while \citet{LAT} explores adversarial attack in the latent space. To the best of our knowledge, no prior work has integrated  adversarial training in VLM safety alignment.

\vspace{-0.5em}
\subsection{Safety of VLMs}
\vspace{-0.5em}
Building upon a backbone LLM, VLMs also face significant safety concerns. To evaluate their safety, several benchmarks \citep{rtvlm, jailbreakv, vlsbench} and jailbreak techniques \citep{figstep, mm-safetybench, visual_adv, imgjp} have been proposed. Jailbreak attacks on VLMs can be categorized into two types: generation-based attacks and optimization-based attacks. Generation-based attacks \citep{figstep, mm-safetybench} create malicious images directly through typography or text-to-image models like Stable Diffusion, while optimization-based attacks \citep{visual_adv, imgjp} distill harmful queries and add imperceptible noise to original images. To address these vulnerabilities, the most prevalent approach is to construct safety-relevant datasets and fine-tune the target model on them. For example, \citet{vlguard} constructs a vision-language safe instruction-following dataset VLGuard and \citet{spa-vl} proposes a safety preference alignment dataset. MMJ-bench \cite{mmj-bench} present a thorough evaluation on existing jailbreak attacks and defenses on various dataset and models. Although these datasets are effective in enhancing the safety of VLMs when facing harmful queries, they do not consider the existence of malicious users.

\vspace{-0.5em}
\section{Methods}
\vspace{-0.5em}
In this section, we introduce \textit{Adversary-aware DPO (ADPO)}. First, we present DPO with \textbf{adversarial-trained reference model} (\textit{AR-DPO}) in section \ref{AR-DPO}, which leverages an adversarially trained model as the reference model for DPO. Then, in section \ref{AT-DPO}, we describe DPO with \textbf{adversarial-aware loss} (\textit{AT-DPO}), which directly incorporates the adversarial min-max optimization framework into the DPO training procedure. Finally, in section \ref{ADPO}, we combine these components to present the \textit{ADPO} framework.

\partitle{Adversarial training} Adversarial training is a min-max optimization framework designed to enhance model robustness against adversarial attacks. It involves two key stages: (1) the adversary generates worst-case perturbations $\delta$
with in certain constrained set $\Delta$ to maximize the model's loss, and (2) the model updates its parameters to minimize the loss on these perturbed inputs. Formally, this can be expressed as:
\begin{equation}
\fontsize{10}{12}
    \min_\theta \max_{\delta \in \Delta} \mathcal{L}(f_\theta(x + \delta), y),
    \vspace{-0.5em}
\end{equation}
where $f_{\theta}$ represents the model, $x$ and $y$ denote the input and output respectively.

\vspace{-0.5em}
\subsection{\textit{AR-DPO}: DPO with Adversarial-trained Reference Model}\label{AR-DPO}
\vspace{-0.5em}

The reference model is the cornerstone of DPO, providing a benchmark to guide the target model's output. However, training the reference model solely under benign conditions without the awareness of the adversarial parties leaves the target model vulnerable to perturbations and susceptible to jailbreak attacks. Therefore, an intuitive approach is to train the reference model with worst-case perturbations, enhancing its resilience to jailbreak attacks and consequently ensuring the target model's robustness.



\partitle{Worst-case perturbation search on image space} 
Since most jailbreak attacks of VLMs are proposed to manipulate image modality, we first consider to search the  worst-case perturbation in the image space. To create a reference model that is aware of the jailbreak attacks in image space, we employ Projected Gradient Descent (PGD) \cite{pgd} to maximize the probability of rejected harmful responses $Y_{r}$. For each harmful image-text pair $x_I$-$x_T$, we optimize the perturbation $\delta$ within a constrained perturbation set $\Delta = \{ \delta \mid x_I+\delta \in [0,1], \left\|\delta\right\|_{p} \le \epsilon \}$. This constraint ensures that each pixel of the perturbed image remains within the valid range, and the maximum perturbation magnitude $\epsilon$ preserves the semantic meaning of the image. The maximization of the probability of rejected responses $Y_{r}$ can be formulated:
\vspace{-0.5em}
{\fontsize{10}{12}
\begin{align}
\label{equation 2}
&\delta^* = \argmax_{\delta \in \Delta} L_{\theta}(x_I, x_T, Y_{r}),\;\text{where}\\
&L_{\theta}(x_I, x_T, Y_{r}) = \log f_{\theta}(Y_{r} \mid x_I+\delta, x_T) 
\vspace{-0.5em}
\end{align}}
This optimization can be solved with Projected Gradient Descent:
\vspace{-0.5em}
\begin{equation}
\fontsize{10}{12}
\delta^{t+1} = \Pi_{\Delta} (x_I^t + \alpha sign\nabla_{x_I^t} L_{\theta}(x_I, x_T, Y_{r})) 
\end{equation}
\vspace{-0.5em}

\partitle{Worst-case perturbation search on latent space} 
To provide a reference model that is also aware of the jailbreak attacks in both text and image domain, we also propose to search for perturbation in the latent space of image-text token embedding. We don't choose to optimize adversarial perturbation over the discrete text token space for two key reasons: (1) optimizing worst-case perturbations in the discrete token space is computationally expensive \cite{harmbench}, and (2) prior studies have shown that such approaches often yield unsatisfactory performance \cite{CAT}.  By operating in the latent space, we achieve a more efficient and effective optimization process in providing an adversarial-aware reference model. Given a VLM $f_\theta$, it can be expressed as the composition of two functions, $f_\theta(Y \mid x_I, x_T) = g_\theta(Y \mid h_\theta(x_I,x_T))$, where $h_\theta$ extracts latent representation of the image-text token embedding, and $g_\theta$ maps these latent activations to the outputs. Similar to the optimization in image space, the search for adversarial perturbation $\delta$ on image-text latent space can be formulated as:
\vspace{-0.5em}
\begin{equation}
\fontsize{10}{12}
\label{equation 3}
    \delta^* = \argmax_{\delta \in \Delta} \log g_\theta(Y_{r} \mid h_{\theta}(x_I, x_T)+ \delta )
    \vspace{-0.5em}
\end{equation}

\partitle{Reference model updates to minimize the loss on perturbed inputs} After generates the worst-case perturbation $\delta^*$, the reference model is adversarially trained to minimize the loss on perturbed inputs. 
The loss is designed to achieve two objectives: (1) maximizing the probability of generating preferred answer on harmful inputs and (2) maintain the general utility on a normal instruction following dataset. To this end, the adversarial training loss consists of two components: the toward loss $\mathcal{L}_{toward}$ to increase the likelihood of preferred safe responses $Y_{p}$  and the utility loss $\mathcal{L}_{utility}$ to preserve the general utility, which can be formulated as: 
\vspace{-0.5em}
\begin{equation}
\fontsize{10}{12}
    \mathcal{L}_{toward} = -\log f_{\theta}(Y_{p}\mid x^{h}_I+\delta^*, x^{h}_T)
\end{equation}
\begin{equation}
\fontsize{10}{12}
    \mathcal{L}_{utility} = -\log f_{\theta}(Y_{util}\mid x^{util}_I, x^{util}_T)
\end{equation}
\vspace{-1em}

If the perturbation is optimized on latent space, the $\mathcal{L}_{toward}$ can be reformulated as:
\vspace{-0.5em}
\begin{equation}
\fontsize{10}{12}
    \mathcal{L}_{toward} = -\log g_\theta(Y_{p} \mid h_{\theta}(x_I^{h}, x_T^{h})+ \delta^* )
    \vspace{-0.5em}
\end{equation}

The overall loss of adversarial training can be formulated as weighted combination of the above two parts and the adversarially trained reference model $f_{\theta_{AT}}$ is optimized with following formula:
\vspace{-0.5em}
\begin{equation}
\fontsize{10}{12}
\label{euqation 7}
    f_{\theta_{AT}} = \argmin_{f_\theta} \mathcal{L}_{toward} + \alpha\mathcal{L}_{utility}
\end{equation}
\vspace{-0.5em}

\partitle{DPO training}
Next, we take the adversarially trained VLM $f_{\theta_{AT}}$ as the reference model for DPO. The objective is to encourage the model to maximize the likelihood of preferred responses while minimizing the likelihood of rejected responses, which can be formulated as:
\vspace{-0.5em}
{\fontsize{10}{12}
\begin{align}
\mathcal{L}_{\text{DPO}} &= -\log \sigma \left( \beta \log \frac{f_{\theta}(Y_{p} | x_I, x_T)}{f_{\theta_{AT}}(Y_{p} | x_I, x_T)} \right. \notag \\
&\quad \left. - \beta \log \frac{f_{\theta}(Y_{r} | x_I, x_T)}{f_{\theta_{AT}}(Y_{r} | x_I, x_T)} \right)
\end{align}}
where $\beta$ is a hyperparameter and controls the penalty the deviations from reference model $f_{\theta_{AT}}$. A higher $\beta$ enforces stricter adherence to the reference model, while a lower
$\beta$ allows more flexibility. The term $\log \frac{f_{\theta}(Y_{p} | x_I, x_T)}{f_{\theta_{AT}}(Y_{p} | x_I, x_T)}$ and $\log \frac{f_{\theta}(Y_{r} | x_I, x_T)}{f_{\theta_{AT}}(Y_{r} | x_I, x_T)}$ measures likelihood of generating the preferred response and rejected answer respectively under the target model $f_{\theta}$ versus the reference model $f_{\theta_{AT}}$. Maximizing the former term encourages the target model to assign higher probability to preferred responses compared to the reference model, while minimizing this term discourages the target model from assigning high probability to rejected responses.

\vspace{-0.5em}
\subsection{\textit{AT-DPO}: DPO Training with Adversarial-aware Loss}
\vspace{-0.5em}
\label{AT-DPO}
Adversarial training can be viewed as the integration of adversarial examples into the training process, and it is independent of the particular choice of the training objective function. Therefore, in addition to utilizing an adversarially trained model as the reference for DPO, we also investigate the potential of direct incorporation of adversarial techniques into the DPO training process. If the perturbation is searched on image space, the loss funtion for \textit{AT-DPO} can be formulated as:
\vspace{-0.5em}
{\fontsize{10}{12}
\begin{align}
\mathcal{L}_{\text{AT-}\text{DPO}} &= -\log \sigma \left( \beta \log \frac{f_{\theta}(Y_{p} | x_I+\delta^* , x_T)}{f_{ref}(Y_{p} | x_I, x_T)} \right. \notag \\
&\quad \left. - \beta \log \frac{f_{\theta}(Y_{r} | x_I+\delta^* , x_T)}{f_{ref}(Y_{r} | x_I, x_T)} \right)
\end{align}
}
where $f_{ref}$ represents a normal reference model without fine-tuning. In each training iteration of DPO, the worst-case perturbation $\delta$ is computed according to Equation \ref{equation 2} and is subsequently added to the input images. 

If the perturbation is optimized on latent space, the loss funtion for \textit{AT-DPO} is:
\vspace{-0.5em}
{\fontsize{10}{12}
\begin{align} 
\mathcal{L}_{\text{AT-}\text{DPO}} &= -\log \sigma \left( \beta \log \frac{g_\theta(Y_{p} \mid h_{\theta}(x_I, x_T)+ \delta^* )}{f_{ref}(Y_{p} | x_I, x_T)} \right. \notag \\
 &\quad \left. - \beta \log \frac{g_\theta(Y_{r} \mid h_{\theta}(x_I, x_T)+ \delta^* )}{f_{ref}(Y_{r} | x_I, x_T)} \right)
\end{align}
}

where $\delta$ is computed according to Equation \ref{equation 3} and then is added to the latent activations.
\vspace{-0.5em}
\subsection{Adversarial-aware DPO (\textit{ADPO})}
\vspace{-0.5em}
\label{ADPO}
Adversarial-aware DPO (\textit{ADPO}) combines both the adversarial reference model and adversarial-aware loss into DPO framework. In  Adversarial reference model training stage, the training procedure follows the adversarial training process of \textit{AR-DPO}, producing a robust and adversarial-aware reference model $f_{\theta_{AT}}$. This model is adversarially trained to generate human-preferred responses under worst-case perturbations, ensuring it serves as a reliable benchmark for the second stage.

In adversarial-aware DPO Training stage, \textit{ADPO} incorporates the adversarial-aware loss of \textit{AT-DPO} directly into the DPO training process. The goal is to optimize the target model $f_{\theta}$while accounting for adversarial conditions. This process can be formulated as:
\vspace{-0.5em}
{\fontsize{10}{12}
\begin{align}
\mathcal{L}_{\text{A-}\text{DPO}} &= -\log \sigma \left( \beta \log \frac{f_{\theta}(Y_{p} | x_I+\delta^* , x_T)}{f_{\theta_{AT}}(Y_{p} | x_I, x_T)} \right. \notag \\
&\quad \left. - \beta \log \frac{f_{\theta}(Y_{r} | x_I+\delta^* , x_T)}{f_{\theta_{AT}}(Y_{r} | x_I, x_T)} \right)
\end{align}
}
\vspace{-1em}

\vspace{-01em}
\section{Experiments}
\vspace{-0.5em}

We begin by detailing our experimental configuration, including the datasets used for \textit{ADPO} training and evaluation, the evaluated jailbreak attacks, and the models tested. Next, we demonstrate the effectiveness of \textit{ADPO} from two perspectives of safety, measured by its robustness against various jailbreak attacks, and utility, evaluated on normal tasks. To further validate our approach, we visualize the shift in latent space, illustrating how \textit{ADPO} enhances robustness. Finally, we conduct an ablation study to support our hyperparameter choices and compare the impact of generating adversarial perturbations in latent space versus image space.

\vspace{-0.5em}
\subsection{Experiment Setup}
\vspace{-0.5em}

\partitle{Safety alignment datasets} Harmful queries can appear in various forms, including adversarial text queries, harmful image-text pairs, or images generated using Stable Diffusion or typographic techniques. To ensure comprehensive safety alignment during fine-tuning, we construct a new dataset based on the HarmBench multimodal (HarmBench-mm) and adversarial training (HarmBench-AT) datasets. Specifically, we sample 80 image-text pairs from HarmBench-mm, pair 40 text samples from HarmBench-AT with blank images, and generate an additional 80 samples using typographic techniques and Stable Diffusion based on HarmBench-AT. This results in a total of 200 harmful image-text pairs. For the utility dataset, we select 500 samples from LLaVA-Instruct-150K to maintain a balance between safety alignment and model utility during fine-tuning.

\partitle{Evaluated VLMs} We evaluate our method on two widely used open-source VLMs: \texttt{LLaVA-1.5-7b}, \texttt{LLaVA-1.6-7b}. We employ LoRA to fine-tune on all linear layers. In our experiments, we specifically evaluate \textit{ADPO} on LLaVA due to its unique capability of converting images into up to 2,880 image tokens. This high tokenization capacity makes LLaVA particularly sensitive to perturbations in the image space. By focusing on LLaVA series, we aim to rigorously test the robustness of \textit{ADPO} under conditions where image perturbations have a pronounced effect, providing a strong benchmark for evaluating the effectiveness of our approach in enhancing adversarial robustness.
Detailed hyperparameters of different fine-tuning setting are provided in Appendix \ref{Hyperparameter}.

\partitle{Evaluated jailbreak attacks} We evaluate two optimization-based attacks, VisualAdv \citep{visual_adv} and MMPGDBlank \citep{harmbench}, on 200 harmful queries from HarmBench standard behaviors. VisualAdv is a universal attack that optimizes a universal adversarial pattern for all harmful behaviors, while MMPGDBlank is a one-to-one attack that optimizes a distinct image for each harmful behavior. Furthermore, we also employ the Jailbreaking subset of MultiTrust \citep{multitrust} to assess the safety of the VLM in a black-box setting. This subset includes three sub-tasks: Typographic Jailbreaking, Multimodal Jailbreaking, and Cross-modal Jailbreaking. Typographic Jailbreaking simply embeds the jailbreaking prompts generated by GPTfuzzer \cite{gptfuzzer} and DAN \cite{dan} into images using typographic methods. Multimodal Jailbreaking involves the random sampling of instances from the existing Multimodal Jailbreak Benchmark \cite{figstep, mm-safetybench}. Cross-modal Jailbreaking investigates whether VLMs are susceptible to adversarial text queries when paired with images, specifically by associating jailbreak prompts with task-relevant images rather than sample-specific images.

\partitle{Evaluated utility benchmark} To evaluate the impact of \textit{ADPO} on normal tasks, we conduct experiments on four widely adopted utilities benchmarks including MMStar \cite{MM-Star}, OCRBench \cite{ocrbench}, MM-Vet \cite{mm-vet}, LLaVABench \cite{llava-1.5}.

\begin{table*}[ht]
    \centering

    \resizebox{\textwidth}{!}{
    \begin{tabular}{c | c c c c c| c c c c }
    \hline
    & \multicolumn{5}{c}{Safety $\downarrow$} & \multicolumn{4}{|c}{Utility$\uparrow$} \\
    \cline{2-10}
    
        &  \multirow{3}{*}{\textbf{VisualAdv}} &  \multirow{3}{*}{\textbf{MMPGDBlank}} & \multicolumn{3}{c|}{\textbf{MultiTrust}} &\multirow{3}{*}{MMStar} &\multirow{3}{*}{OCRBench} & \multirow{3}{*}{MM-Vet} & \multirow{3}{*}{LLaVABench} \\
          \cline{4-6}
         
   &  & & Typographic& Multimodal & Crossmodal \\
  &  & & Jailbreak & Jailbreak & Jailbreak \\   
\hline
     LLaVA-1.5-7b  & 64.5 & 84.0 & 22.2 & 55.1 & 42.0 & 32.7 & 202  & 29.9 & 59.5\\

    +Supervised FT& 19.0 & 76.0 & 0.5 & 10.3 & 27.1 & 33.7 (\textcolor{red}{$\uparrow$}) & 201  & 28.6 & 53.6\\

    + DPO & 12.0 & 33.0 & 0.7 & 8.8 & 9.6& 33.9 (\textcolor{red}{$\uparrow$}) & 198  & 28.9 & 54.4 \\

    +\textit{AR-DPO} & \colorbox{gray!30}{\textbf{2.5}} & 1.0 & \colorbox{gray!30}{\textbf{0.0}} & \colorbox{gray!30}{\textbf{0.0}} & 2.4& \underline{34.1} (\textcolor{red}{$\uparrow$})& 187  & 23.3 & 47.7 \\

    +\textit{AT-DPO} & 7.5 & 8.5 & 0.5 & 3.4 & 9.1 & 33.4 (\textcolor{red}{$\uparrow$})& \underline{193}  & \underline{28.9} & \underline{51.6} \\

    + \textit{ADPO} & 5.0 & \colorbox{gray!30}{\textbf{0.5}} & \colorbox{gray!30}{\textbf{0.0}} & \colorbox{gray!30}{\textbf{0.0}} & \colorbox{gray!30}{\textbf{0.2}} &33.7 (\textcolor{red}{$\uparrow$})& 184& 24.2 & 48.2\\ 
    \hline

     LLaVA-1.6-7b  & 33.5 & 48.5 & 8.5 & 58.3 & 56.2 & 37.9 & 500  & 43.1 & 66.8\\

    +Supervised FT& 6.5  & 22.5 & 2.0 & 25.4 & 34.2  & 38.2 & 501 (\textcolor{red}{$\uparrow$}) & 40.0 & 58.6\\

    + DPO & 2.0 & 7.0 & 1.2 & 7.1 & 27.1 & 38.1  (\textcolor{red}{$\uparrow$}) & 489  & 38.3 & 59.1\\

    +\textit{AR-DPO} & \colorbox{gray!30}{\textbf{0.0}} & 8.5 & 0.2 & \colorbox{gray!30}{\textbf{0.0}} & \colorbox{gray!30}{\textbf{2.4}} & \underline{37.7} & 436  & 38.0 & 50.5\\

    +\textit{AT-DPO} & 0.5 & 3.5 & 0.5 & 4.9 & 21.3 & 36.9  & \underline{448}& \underline{38.9} & \underline{58.2}\\

    + \textit{ADPO} & \colorbox{gray!30}{\textbf{0.0}} & \colorbox{gray!30}{\textbf{0.0}} & \colorbox{gray!30}{\textbf{0.0}} & 0.2 & 8.4 & 36.9 & 433 & 37.6 & 50.9 \\
    \hline
\end{tabular}}
\vspace{-1em}
\caption{Safety and utility evaluation of \textit{ADPO}, its ablations, and baselines on \texttt{LLaVA-1.5} and \texttt{LLaVA-1.6}. For safety evaluation, the lowest ASR for each jailbreak attack is highlighted in bold and gray shadow. For utility evaluation, the highest score among \textit{ADPO} and its ablations is underlined. Cases where the utility score improves after safety alignment compared to the original model are marked with \textcolor{red}{$\uparrow$}.}
\label{safety evaluation}
\vspace{-1em}
\end{table*}

\vspace{-0.5em}
\subsection{Safety Evaluation}
\vspace{-0.5em}
In this section, we evaluate the effectiveness of \textit{ADPO} in improving safety alignment. We compare \textit{ADPO} against baselines including supervised fine-tuning (SFT) and standard DPO, as well as its ablations: \textit{AR-DPO} (adversarial-trained reference model only) and \textit{AT-DPO} (adversarial-aware DPO loss only).  The evaluation focuses on Attack Success Rate (ASR) across various jailbreak attacks, which is defined as the fraction of successful attacks over all tested examples. The HarmBench classifier \cite{harmbench} is employed to determine whether the model responses are harmful. 

As shown in the safety column of Table \ref{safety evaluation},  \textit{ADPO} and its ablations (\textit{AR-DPO} and \textit{AT-DPO}) significantly reduce the ASR across all jailbreak attacks on both \texttt{LLaVA-1.5} and \texttt{LLaVA-1.6}, outperforming SFT and standard DPO.  Specifically, \textit{ADPO} emerges as the most effective method, reducing the ASR to nearly 0 across almost all attacks, underscoring the importance of integrating both the adversarial aware-reference model and adversarial-aware DPO loss.  

In addition, we can notice that \textit{AT-DPO} is not very effective on Crossmodal jailbreak, compared with \textit{AR-DPO} and \textit{ADPO}, highlighting the importance of including the adversarial-aware reference model. The Crossmodal Jailbreaking dataset consists of text-level jailbreak prompts. Since \textit{AT-DPO} adds perturbation only to the image space, it may not generalize well to text-level attacks through a single-stage safety alignment. In contrast, \textit{AR-DPO} and \textit{ADPO}, which utilize an adversarial trained model as reference model, demonstrate a greater ability to recognize harmful semantics in a harmful query, even when the harmfulness originates from text inputs. Although SFT and DPO exhibit comparable performance on some cases in the Multitrust benchmark, they demonstrate reduced effectiveness against white-box optimization-based attacks. Notably, the MMPGDBlank attack maintains a high ASR, with values of 33.0 and 7.0 for DPO, and 76.0 and 22.5 for FT on \texttt{LLaVA-1.5} and \texttt{LLaVA-1.6} respectively. In contrast, \textit{ADPO} achieved 0.5 and 0 ASR on MMPGDBlank.


\begin{figure}[t]
    \centering
    \includegraphics[width=.4\textwidth]{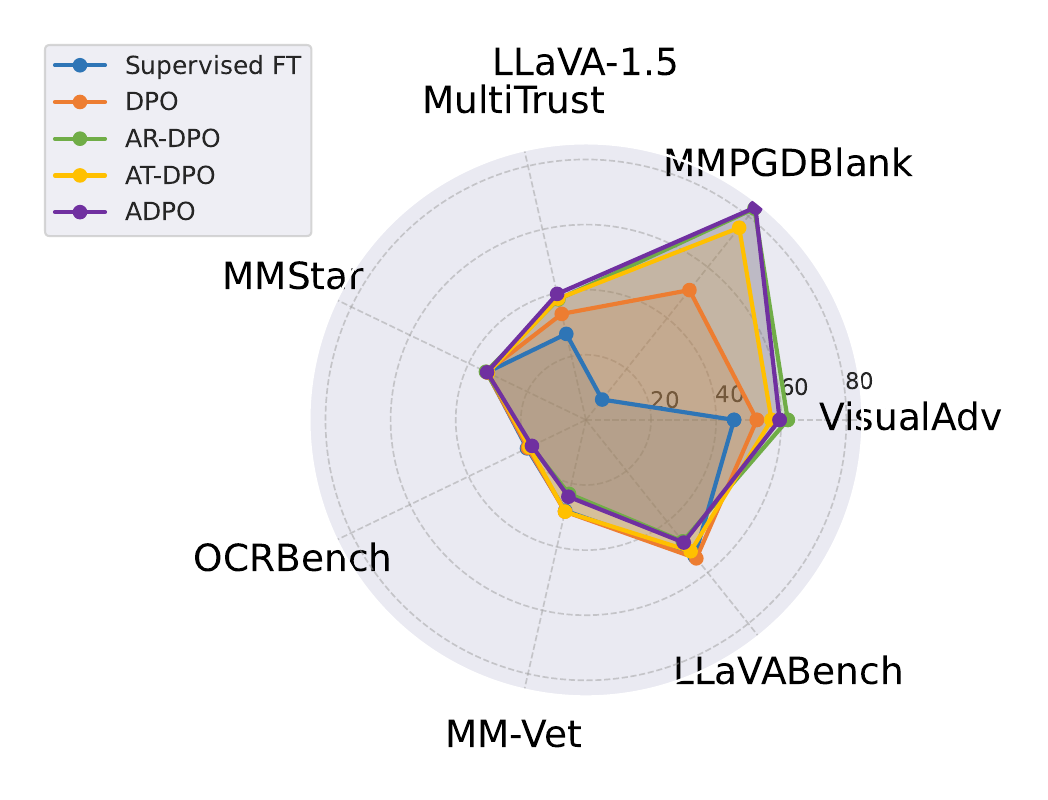}
    \setlength{\abovecaptionskip}{0.2cm}
    \vspace{-1em}
    \caption{Safety-utility trade-off, where jailbreak dimensions indicate the ASR reduction (the larger the better). A larger area for each method represents more effective in safety alignment and utility maintainness. } 
    \label{radar}
    \vspace{-1em}
\end{figure}

\vspace{-0.5em}
\subsection{Utility Evaluation}
\vspace{-0.5em}
\textit{ADPO}, along with its ablations and baselines is evaluated on four normal task benchmarks, each has its own evaluation metric (detailed in Appendix \ref{utility benchmark}). MMStar focuses on image-based multiple-choice questions, while the other three benchmarks are visual question answering (VQA) datasets. The results are shown in the utility column of Table \ref{safety evaluation}. For all datasets, a higher score indicates better performance on that dataset. The highest score among \textit{ADPO} and its ablations is underlined. Cases where the utility score improves after safety alignment compared to the original model are marked with \textcolor{red}{$\uparrow$}.

Overall, all methods somehow reduce the utility score on VQA bechmarks, whihe  multiple-choice dataset MMStar experience an increase in the utility score after safety fine-tuning, indicating its less sensitive to the safety alignment.
Although \textit{ADPO} and \textit{AR-DPO} demonstrate remarkable performance in enhancing robustness against jailbreak attacks, we observe a slight trade-off on the VQA datasets. This indicates that the adversarial training process, while enhancing safety, may inadvertently lead to a more conservative model behavior, occasionally affecting its ability to handle benign queries. This finding suggests the necessity to explore refined fine-tuning strategies and objective functions in the future work to further optimize this balance.

\begin{figure*}[h]
    \centering
        
    \begin{subfigure}[b]{1\textwidth}
        \centering
        \begin{minipage}{0.01\textwidth}
        \centering
        \rotatebox{90}{\tiny  \textbf{MMPGDBlank}}
    \end{minipage}
    \begin{minipage}{0.95\textwidth}
        \includegraphics[width=\linewidth]{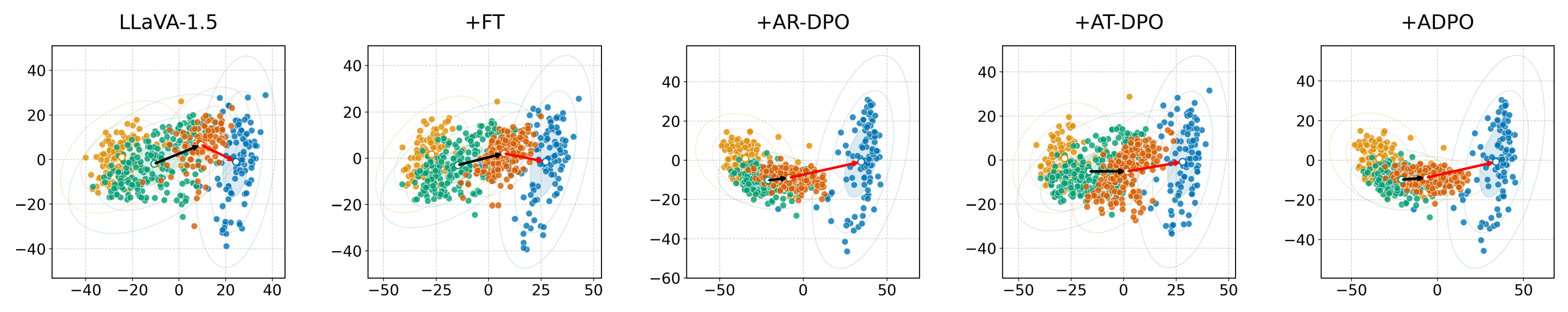}
        \label{figvis:sub1}
        \end{minipage}
    \end{subfigure}

    \vfill
    \vspace{-1.5em}

    \begin{subfigure}[b]{1\textwidth}
        \centering
                \begin{minipage}{0.01\textwidth}
        \centering
        \rotatebox{90}{\tiny  \textbf{VisualAdv}}
    \end{minipage}
        \begin{minipage}{0.95\textwidth}
        \includegraphics[width=\linewidth]{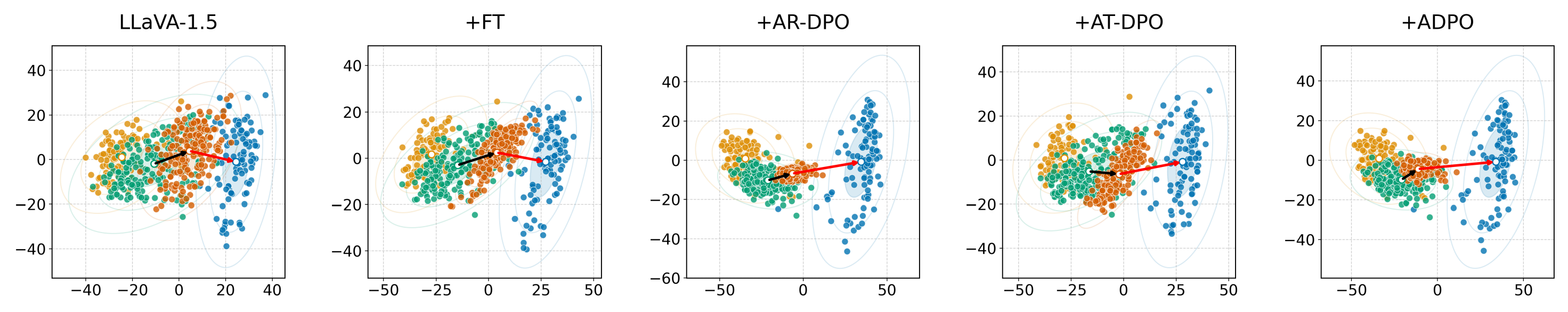}
        \label{figvis:sub2}
        \end{minipage}
    \end{subfigure}

\centerline{
\textcolor[rgb]{0.8705882352941177, 0.5607843137254902, 0.0196078431372549}{\small$\bullet$ Harmful anchor query} \textcolor[rgb]{0.00392156862745098, 0.45098039215686275, 0.6980392156862745}{\small$\bullet$ Harmless anchor query} \textcolor[rgb]{0.00784313725490196, 0.6196078431372549, 0.45098039215686275}{\small$\bullet$ HarmBench query} \textcolor[rgb]{0.8352941176470589, 0.3686274509803922, 0.0}{\small$\bullet$ HarmBench query +  Attack}}
    \caption{Visualization of representation space of \texttt{LLaVA-1.5} trained with \textit{ADPO}, its ablations and FT. (1) Harmbench queries (green) are closer to the harmful anchor cluster (yellow) , demonstrating the model's success in recognizing their harmfulness. (2) \texttt{LLaVA-1.5} trained with \textit{ADPO} and its ablations successfully moves the orange cluster closer to the harmful (yellow) and HarmBench (green) clusters (black arrow) while pushing it further from the harmless cluster (blue, red arrow), indicates that the safety aligned model can better recognize the harmfulness in  Harmbench queries even with the existence of jailbreak attacks.}
    \label{visualization}
    \vspace{-1em}
\end{figure*}

\partitle{Safety and utility trade-off}
To further evaluate the safety-utility trade-off, we present a radar chart in Figure \ref{radar}. Note that the jailbreak dimensions indicate the ASR reduction (the larger the better) and MultiTrust dimension denotes the average ASR reduction across its sub-tasks.  A larger area represents more effective in safety alignment and utility maintainess.  As shown in Figure \ref{radar}, the area for \textit{ADPO} (purple area) and  \textit{AR-DPO} (green are) are the largest compared with SFT and DPO.


\vspace{-0.5em}
\subsection{Latent Space Representation Analysis}
\vspace{-0.5em}
Shifts in the latent space representation of harmful queries towards the the direction of harmless query can reveal the mechanisms of jailbreak attacks \cite{lin2024towards}. 

Similarly, to further validate the effectiveness of \textit{ADPO}, we visualize the representation space of \texttt{LLaVA-1.5} using the output of LLM's last hidden state, which captures comprehensive information from the entire sequence. Specifically, we employ principle component analysis (PCA) \cite{PCA} to analysis four types of queries: Harmful anchor query, Harmless anchor query, HarmBench query, HarmBench query with Attack. The harmful and harmless anchor queries, collected from  \cite{zheng2024prompt}, serve as reference points for general harmful and harmless queries, exhibiting significant differences in harmfulness while maintaining similar query formats and text lengths.


As shown in Figure \ref{visualization}, the representations of harmful and harmless anchor queries form distinct clusters (yellow and blue), indicating the model's ability to differentiate between harmful and harmless semantics. Harmbench queries, which is indicated as green clusters are closer to the harmful anchor cluster (yellow), demonstrating the model's success in recognizing their harmfulness. However, after jailbreak attacks (MMPGDBlank and VisualAdv), HarmBench queries shift significantly towards the harmless cluster (blue), as seen in the orange clusters in the first column of Figure \ref{visualization}.

We compare the latent space of \texttt{LLaVA-1.5} trained with \textit{AR-DPO}, \textit{AT-DPO}, \textit{ADPO} and SFT in the subsequent columns of Figure \ref{visualization}. Notably, \texttt{LLaVA-1.5} trained with \textit{ADPO} and its ablations successfully moves the orange cluster closer to the harmful (yellow) and HarmBench (green) clusters (black arrow) while pushing it further from the harmless cluster (blue, red arrow). In contrast, the SFT model fails to exhibit this behavior. This finding indicates that the safety aligned model can better recognize the harmfulness in  Harmbench queries even with the existence of jailbreak attacks.  



\vspace{-.5em}
\subsection{Ablation Study}
\vspace{-.5em}

\begin{figure}[ht]
\setlength{\abovecaptionskip}{0.cm}
    \centering
    \begin{subfigure}[b]{0.225\textwidth}
        \centering
        \includegraphics[width=\linewidth]{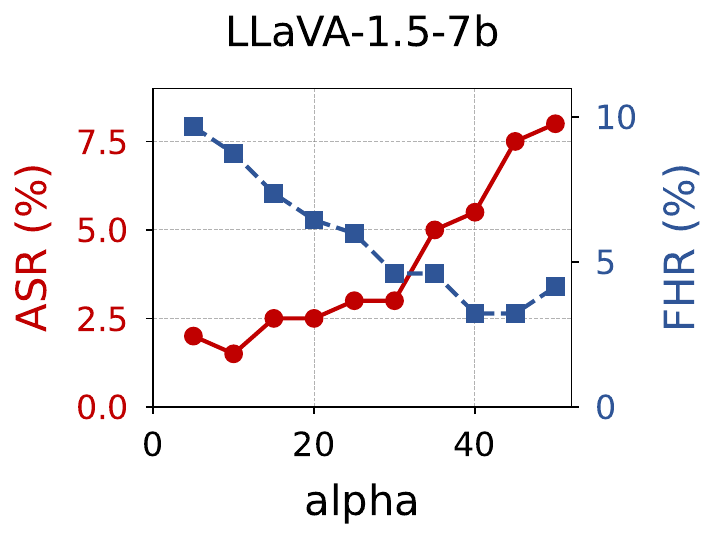}
        \label{fig:sub1}
    \end{subfigure}
    \hfill
    \begin{subfigure}[b]{0.225\textwidth}
        \centering
        \includegraphics[width=\linewidth]{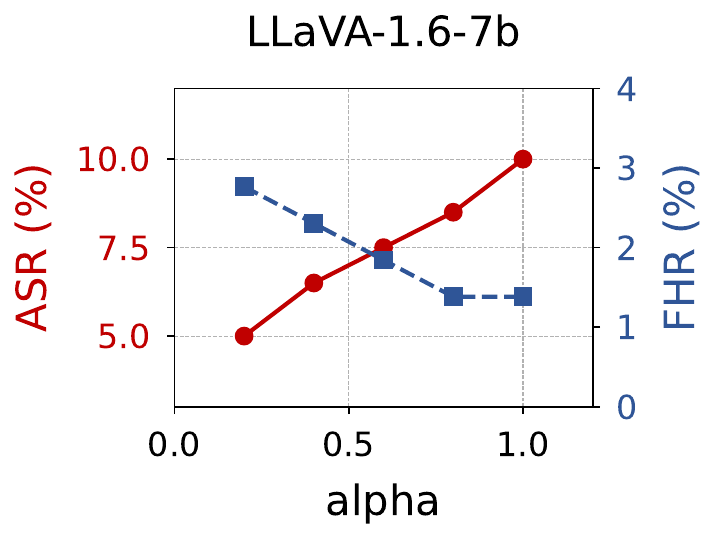}
        \label{fig:sub2}
    \end{subfigure}
\vspace{-1em}
    \caption{Ablation study on adversarial training $\alpha$.}
    \label{ablation study}
\vspace{-1em}
\end{figure}

Figure \ref{ablation study} presents an ablation study on $\alpha$ in Equation \ref{euqation 7}, which balance the trade-off between safety and utility during adversarial training. The left Y-axis displays the ASR, while the right Y-axis illustrates the False Harm Rate (FHR) on MM-Vet, representing the proportion of benign samples incorrectly flagged as harmful. The optimal goal is to minimize both ASR (enhancing safety robustness) and FHR (preserving utility). Based on the intersection of the two curves, we select the appropriate $\alpha$ value for our experiments.

\vspace{-.5em}
\subsection{Latent Space Adversarial Training}
\vspace{-.5em}

We also investigate the search of adversarial perturbations in the latent space of image-text embeddings, introduced in Section \ref{AR-DPO}. Specifically, we perform adversarial perturbations at layers 8, 16, 24, and 30 of the backbone LLM for the VLM.  As shown in Table \ref{lat}, where \textit{L-ADPO}, \textit{L-AR-DPO} and \textit{L-AT-DPO} represent the latent space counterparts of ADPO and its ablations. The results indicate that both \textit{L-AR-DPO} and \textit{L-ADPO} exhibit similar performance with their counterparts in the image space. However, \textit{L-AT-DPO} yields a slightly negative result compared with \textit{AT-DPO}. This suggests that adversarial training in the latent space may lead to overfitting to particular adversarial patterns within the latent space, potentially hindering its generalization to natural harmful queries.

\begin{table}[h]
    \centering

    \resizebox{\columnwidth}{!}{
    \begin{tabular}{c |c c c c| c }
    \hline
    & \multicolumn{4}{c}{Safety $\downarrow$} & \multicolumn{1}{|c}{Utility$\uparrow$} \\
    \cline{2-6}
    &  \multirow{2}{*}{\textbf{MMPGDBlank}} & \multicolumn{3}{c|}{\textbf{MultiTrust}} & \multirow{2}{*}{MM-Vet}  \\
         
     & & Typo& Multimodal & Cross \\
\hline
     LLaVA-1.5-7b  & 84.0 & 22.2 & 55.1 & 42.0 & 29.9 \\[2pt]
\hline
    +\textit{AR-DPO} & 1.0 & 0.0 & 0.0 & 2.4 & 23.3 \\[2pt]

    +\textit{AT-DPO} & 8.5 & 0.5 & 3.4 & 9.1 & 28.9 \\[2pt]

    + \textit{ADPO} & 0.5& 0.0 & 0.0 & 0.2 & 24.2 \\[2pt]
    \hline

    +\textit{L-AR-DPO} & 2.5 & 0.0 & 0.0 & 1.6 & 23.4\\[2pt]

    +\textit{L-AT-DPO} & 31.5 & 1.0 & 23.1 & 14.9 & 28.9\\[2pt]

    + \textit{L-ADPO} & 2.0 & 0.0 & 0.0 & 2.2  & 25.1  \\[2pt]
    \hline

\end{tabular}}
\vspace{-1em}
\caption{Comparison of worst-case perturbation searched in the image space versus in the latent space of image-text embedding.}
    \label{lat}
    \vspace{-1em}
\end{table}

\vspace{-0.5em}
\section{Conclusion}
\vspace{-0.5em}
We propose \textit{ADPO}, a novel training framework to enhance safety alignment of Vision-Language Models (VLMs) under adversarial scenarios. Compared with baselines, \textit{ADPO} demonstrates its effectiveness through extensive experiments, achieving an ASR close to 0 across nearly all jailbreak attacks. Furthermore, we also visualize the shift in the latent space to further validate the effectiveness of \textit{ADPO}. The results underscore the potential of \textit{ADPO} as a robust solution for enhancing the safety alignment of VLMs. It would be interesting to investigate refined fine-tuning strategies that better balance the trade-off between safety and utility in the future.

\section*{Limitations}
We outline the limitations of our study as follows: 

1. While enhancing the safety robustness of VLMs, \textit{ADPO} can inevitably compromise their general performance on utility benchmarks, underscoring the need for better optimization of this trade-off in future research. 

2. We only focus on integrating adversarial training into the training process of DPO. The exploration of incorporating adversarial training into other alignment algorithms, such as RLHF or IPO \cite{IPO}, is reserved for future work.

3. In this study, we focus solely on using PGD as the method for generating adversarial perturbations. Therefore, it is worthy to investigate the adaptation of other adversarial attacks, such as C\&W attack \cite{carlini2017towards}, to optimize adversarial perturbations.

\section*{Ethics Statements}
In this paper, we propose a safety alignment framework to enhance the safety robustness of VLMs against jailbreak attacks. We believe that the adoption of \textit{ADPO} will significantly contribute to the development of more secure and robust VLMs in the future, enhancing their safety and reliability in a wide range of applications. We acknowledge that data utilized for training and evaluation in our paper may contain harmful content and is strictly limited to the model training and evaluation process. \textit{ADPO} training framework will be released in the near future and contributes to the advancement of safer VLMs.

\bibliography{acl_latex}

\cleardoublepage
\appendix

\section{Utility Benchmarks}
\label{utility benchmark}
\partitle{MMStar} MMStar is a benchmark for multimodal multiple-choice questions, consisting of 1,500 samples that assess six fundamental capabilities of vision-language models (VLMs): fine-grained perception, coarse perception, mathematics, science and technology, logical reasoning, and instance reasoning. The metric used to evaluate MMStar is accuracy and is calculated by some heuristic rules.

\partitle{OCRBench} OCRBench is a comprehensive Optical Character Recognition (OCR) benchmark to assess the OCR capabilities for VLMs. It comprises 1,000 question-answer pairs, and its evaluation metric is based on the number of outputs that match the ground truth answers.

\partitle{MM-Vet} MM-Vet is an evaluation benchmark that examines VLM on six core capabilities, including recognition, OCR, knowledge, language generation, spatial awareness, and math. For each sample, MM-Vet score is calculated by GPT-4 based on the input question, ground truth, and model output.

\partitle{LLaVABench} LLaVABench contains 60 samples in three categories: conversation, detailed description, and complex reasoning. The evaluation score is determined by GPT-4, which compares the generated answer to a reference answer.

\section{Hyperparameter Choices}
\label{Hyperparameter}

Table \ref{Hyperparameter table} presents a full list of hyperparameter choices for each fine tuning method. 

\begin{table}[ht]
    \centering
    \resizebox{\columnwidth}{!}{
    \begin{tabular}{c c|c c c c c c}
    \hline
       & Hyperparameter & FT & AT & DPO & \textit{AR-DPO} & \textit{AT-DPO} & \textit{ADPO} \\
        \hline
       \multirow{7}{*}{\rotatebox{90}{LLaVA-1.5-7b}} &Learning rate & 2e-5 & 2e-5 & 2e-5 & 2e-5 & 2e-5 & 2e-5 \\
        &Batch size & 64 & 64 & 64 & 64 & 64 & 64 \\
        &Epochs & 2 & 2 & 10 & 5 & 10 & 5 \\
        &$\alpha$ & 30 & 30  & - & - & - & -  \\
        &$\beta$ & - & - & 0.1 & 0.01 & 0.1 & 0.01 \\
        &Lora r & 128 & 128 & 128 & 128 & 128 & 128 \\
        &Lora alpha & 256 & 256 & 256 & 256 & 256 & 256 \\
        \hline
       \multirow{7}{*}{\rotatebox{90}{LLaVA-1.6-7b}} &Learning rate & 2e-5 & 2e-5 & 2e-5 & 2e-5 & 2e-5 & 2e-5 \\
        &Batch size & 64 & 64 & 64 & 64 & 64 & 64 \\
        &Epochs & 2 & 2 & 10 & 5 & 10 & 5 \\
        &$\alpha$ & 0.6 & 0.6  & - & - & - & -  \\
        &$\beta$ & - & - & 0.1 & 0.1 & 0.1 & 0.1 \\
        &Lora r & 128 & 128 & 128 & 128 & 128 & 128 \\
        &Lora alpha & 256 & 256 & 256 & 256 & 256 & 256 \\
        \hline        
    \end{tabular}}
    \caption{Hyperparameters for \texttt{LLaVA-1.5-7b} and \texttt{LLaVA-1.6-7b} with different fine-tuning settings.}
\label{Hyperparameter table}
\end{table}

\section{Additional Experimental Results}
\subsection{Radar chart of LLaVA-1.6}
\begin{figure}[t]
    \centering
    \includegraphics[width=.45\textwidth]{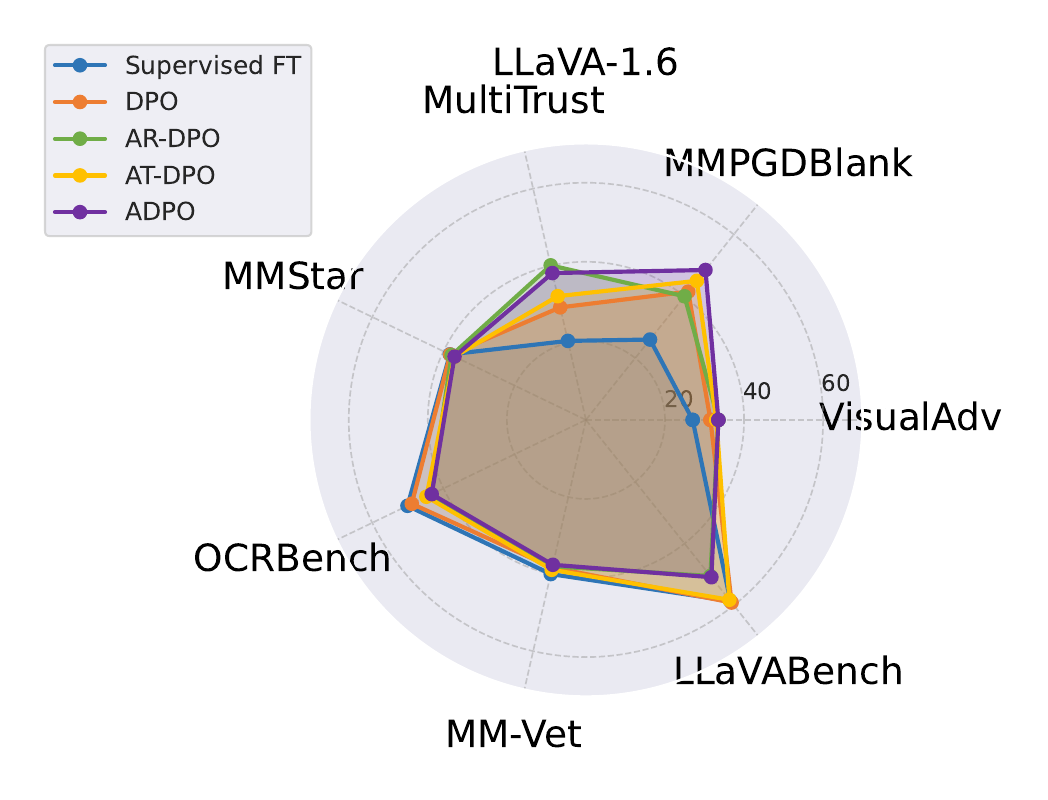}
    \setlength{\abovecaptionskip}{0.2cm}
    \caption{This graph illustrates the reduction in ASR and utility score of \textit{ADPO}, its ablations and baselines over different jailbreak attacks and utility benchmarks on \texttt{LLaVA-1.6}.} 
    \label{radar_16}
\end{figure}

The radar chart of \texttt{LLaVA-1.6} are presented in Figure \ref{radar_16}.

\subsection{Latent Space Adversarial Training on LLaVA-1.6}

The comparision of adversarial training on latent sapce versus image space on \texttt{LLaVA-1.6} are shown in Tabel \ref{latent adversarial training on LLaVA-1.6}.

\begin{table}[h]
    \centering

    \resizebox{\columnwidth}{!}{
    \begin{tabular}{c |c c c c| c }
    \hline
    & \multicolumn{4}{c}{Safety $\downarrow$} & \multicolumn{1}{|c}{Utility$\uparrow$} \\
    \cline{2-6}
    
         &  \multirow{2}{*}{\textbf{MMPGDBlank}} & \multicolumn{3}{c|}{\textbf{MultiTrust}} & \multirow{2}{*}{MM-Vet}  \\
          \cline{3-5}
         
     & & Typo& Multimodal & Cross \\
\hline
     LLaVA-1.6-7b  & 48.5 & 8.5 & 58.3 & 56.2 & 43.1 \\[2pt]
\hline
    +\textit{AR-DPO} & 8.5 & 0.2 & 0.0 & 2.4 & 38.0 \\[2pt]

    +\textit{AT-DPO} & 3.5 & 0.5 & 4.9 & 21.3 & 38.9 \\[2pt]

    + \textit{ADPO} & 0.5& 0.0 & 0.2 & 8.4 & 37.6 \\[2pt]
    \hline

    +\textit{L-AR-DPO} & 11.0 & 1.0 & 0.0 & 21.6 & 41.0\\[2pt]

    +\textit{L-AT-DPO} & 12.0 & 1.7 & 8.5 & 29.1 & 39.6\\[2pt]

    + \textit{L-ADPO} & 10.5 & 1.2 & 0.0 & 24.9  & 42.6  \\[2pt]
    \hline

\end{tabular}}

\caption{Comparison of worst-case perturbation searched in the image space versus in the latent space of image-text embedding on \texttt{LLaVA-1.6}.}
    \label{latent adversarial training on LLaVA-1.6}
\end{table}

\section{Computing resources}
The experiments are carried by 2*NVIDIA A40 gpus. All conducted experiments required at least 768 gpu hours.

\section{AI Assistants}
We only used AI for grammar correction and sentence polishing in the paper.

\end{document}